\documentclass{elsart}

\usepackage{graphicx,amssymb}
\DeclareGraphicsExtensions{.eps}
\usepackage[authoryear]{natbib}

\journal{Planetary and Space Science}

\begin{document}

\begin{frontmatter}

\title{Variability of the methane trapping in martian subsurface clathrate hydrates}

\author{Caroline Thomas},
\ead{caroline.thomas@univ-fcomte.fr}
\author{Olivier Mousis},
\author{Sylvain Picaud},
\and
\author{Vincent Ballenegger}

\address{Universit{\'e} de Franche-Comt{\'e}, Institut UTINAM, CNRS/INSU, UMR 6213, \\
25030 Besan\c{c}on Cedex, France}

\begin{abstract}
Recent observations have evidenced traces of methane (CH$_4$) heterogeneously distributed in the martian atmosphere. However, because the lifetime of CH$_4$ in the atmosphere of Mars is estimated to be around 300--600 years on the basis of photochemistry, its release from a subsurface reservoir or an active primary source of methane have been invoked in the recent literature. Among the existing scenarios, it has been proposed that clathrate hydrates located in the near subsurface of Mars could be at the origin of the small quantities of the detected CH$_4$. Here, we accurately determine the composition of these clathrate hydrates, as a function of  temperature and gas phase composition, by using a hybrid statistical thermodynamic model based on experimental data. Compared to other recent works, our model allows us to calculate the composition of clathrate hydrates formed from a more plausible composition of the martian atmosphere by considering its main compounds, i.e. carbon dioxyde, nitrogen and argon, together with methane. Besides, because there is no low temperature restriction in our model, we are able to determine the composition of clathrate hydrates formed at temperatures corresponding to the extreme ones measured in the polar caps. Our results show that methane enriched clathrate hydrates could be stable in the subsurface of Mars only if a primitive CH$_4$-rich atmosphere has existed or if a subsurface source of CH$_4$ has been (or is still) present.
\end{abstract}


\date{\today}

\end{frontmatter}

%\linenumbers

\section{Introduction}

The Planetary Fourier Spectrometer (PFS) onboard the {\it Mars
Express} spacecraft has detected $\sim$ 10 parts per billion by
volume (ppbv) of methane in the atmosphere of Mars (Formisano et
al., 2004). Comparable abundances have been reported by Mumma et
al. (2003), by Krasnopolsky et al. (2004), and very
recently by Villanueva et al. (2008) from ground-based
observations. Because the photochemical mean lifetime of the
atmospheric methane is about 300--600 years (Krasnopolsky et al.,
2004; Formisano et al., 2004), its release from a subsurface
reservoir or an active primary source of methane have been invoked
in the recent literature. It has thus been suggested that the
methane observed on Mars could be of biogenic origin and would
originate from organisms living in the near subsurface (Formisano
et al., 2004; Krasnopolsky et al., 2004; Krasnopolsky, 2006) or
that the olivine hydratation in the martian regolith or crust
could be a major methane source (Oze and Sharma, 2005).
Alternatively, it has been proposed that the methane observed on
Mars could be produced from photolysis of water in the presence of
CO (Bar-Nun and Dimitrov, 2006).

The atmospheric methane of Mars could also be tied to the possible presence of methane clathrate hydrates that decompose in the near-subsurface (Prieto-Ballesteros et al., 2006; Chastain and Chevrier, 2007). In this case, clathrate hydrates
don't bring any constraint on the origin of the martian methane.
Indeed, their ability to retain methane over long timescales make
them a secondary reservoir which can be filled either by ancient
or by current methane sources (Prieto-Ballesteros et al., 2006;
Chastain and Chevrier, 2007). In this context, the stability of
methane clathrate hydrates under martian conditions has been
explored by Chastain and Chevrier (2007) (hereafter CC07) with the
use of the program CSMHYD developed by Sloan (1998). From these
calculations, CC07 determined the plausible composition of methane
clathrate hydrates that could be located in the martian crust or
in the polar caps. They also discussed the possible sources that
may be at the origin of these clathrate hydrates.

However, these  authors only considered the formation of binary
CO$_2$-CH$_4$ clathrate hydrates from the martian atmosphere, due
to the limited list of available molecules in the CSMHYD program.
With a fractional abundance of 95\%, carbon dioxyde is the main
atmospheric compound, but other molecules such as nitrogen and
argon, whose abundances reach $\sim$ 2.7 and 1.6 \% respectively
(Moroz, 1998), have been neglected in the composition and
stability calculations of clathrate hydrates by CC07. As shown by
Thomas et al. (2007, 2008) in the case of Titan, minor compounds
can strongly affect the composition of clathrate hydrates formed
from the atmospheric gas phase and also their temperature
dependence. Moreover, because the CSMHYD code is not suitable
below $\sim$ 171 K for CO$_2$-CH$_4$ gas mixtures relevant to the
atmospheric composition of Mars, CC07 did not calculate the
composition of clathrate hydrates possibly formed in the polar
caps where the {\it in situ} temperature can decrease
down to $\sim$ 130 K (Kieffer et al., 2001).

In the present work, we reinvestigate the assumptions of CC07 and
determine more accurately the composition of clathrate hydrates
that may form in the near subsurface of Mars as a function
of temperature and gas phase composition, by using a statistical
thermodynamic model based on experimental data and on the original
work of van der Waals and Platteeuw (1959). Compared to the work of CC07, our model allows us to
calculate the composition of clathrate hydrates formed from a more
plausible composition of the martian atmosphere by considering its
main compounds, i.e. carbon dioxyde, nitrogen and argon, together
with methane. Besides, because there is no low temperature
restriction in our model, we are able to determine the composition
of clathrate hydrates formed at temperatures corresponding to the
extreme ones measured in the polar caps.

\section{Theoretical background}
\subsection{Model}

To calculate the relative abundance of CH$_4$ incorporated in
clathrate hydrates on Mars, we used the same approach as in our
previous studies devoted to the trapping of noble gases by
clathrate hydrates on Titan (Thomas et al., 2007, 2008).
This approach is based on the statistical model developed
by van der Waals and Platteuw (1959) and it is similar to the one
proposed by Sloan (1998) in the CSMHYD program. However, it
differs from this latter approach by the use of experimentally
determined dissociation curves in our code instead of calculated
dissociation pressures (Thomas et al., 2008). This allows us to
determine the relative abundances in clathrate hydrates down to
very low temperatures whereas the CSMHYD program is for example
limited to $\sim$ 171 K for the clathrate hydrates considered
here.

\begin{figure}
\resizebox{\hsize}{!}{\includegraphics[angle=0]{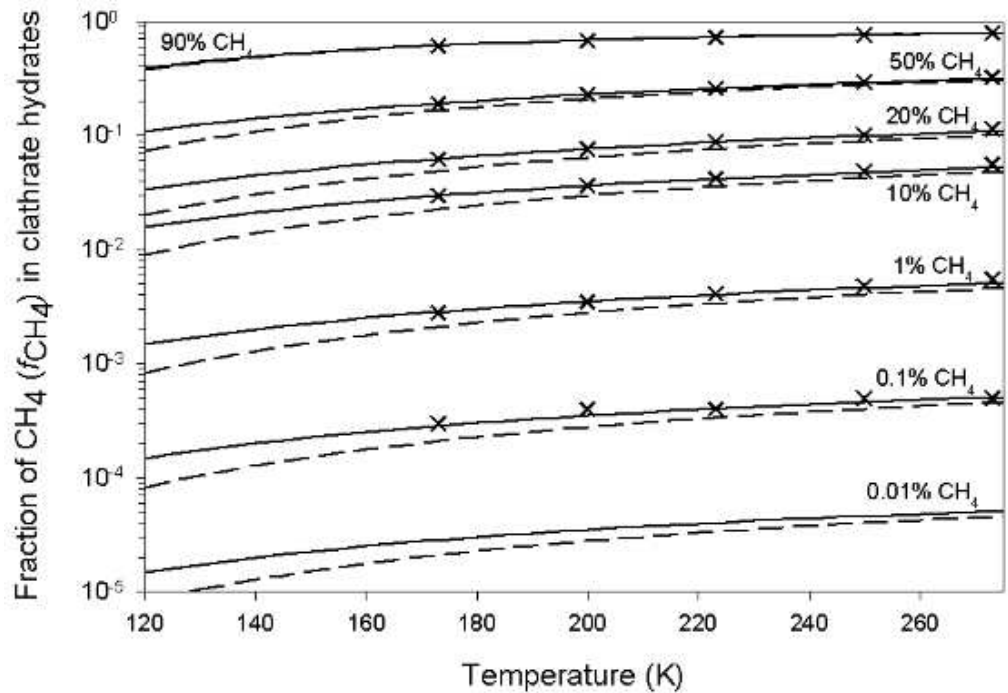}}
\caption{Fraction of CH$_4$ trapped in clathrate hydrates as a function of temperature and of the different CH$_4$/CO$_2$ mixture ratios adopted in the gas phase (see text). The calculations have been performed using either the CSMHYD program proposed by Sloan (1998) (crosses) or the approach proposed in the present paper (solid and dashed lines). The solid lines represent the results obtained with the parameters of the Kihara potential from Sloan (1998). The dashed lines represent the results obtained with the parameters of the Kihara potential from Parrish and Prausnitz (1972) for all species, except for the parameter $a$ for CO$_2$, which comes from Sloan (1998).}
\label{fig1}
\end{figure}

In order to validate the calculations performed in the present
study, we have compared the results issued from our approach to those
obtained using the CSMHYD program when considering gas phases
containing CH$_4$ and CO$_2$ in different initial abundances. As
an illustration, we present on Figure \ref{fig1} the relative
abundances of methane in mixed CO$_2$/CH$_4$ clathrate hydrates as
calculated with our code by considering several initial abundances
of CH$_4$ in the gas phase (0.01\%, 0.1\%, 1\%, 10\%, 20\%, 50\%
or 90\% (the relative abundance of CO$_2$ in each case is
determined such as $x_{\rm CH_4}+x_{\rm CO_2}=1$). Figure
\ref{fig1} clearly shows that CH$_4$ abundances in clathrate hydrates
obtained from our program (full line) using the Kihara parameters
and cage geometries given by Sloan (1998) fit very well those
obtained with the CSMHYD program (crosses) down to 171 K, that is
in the range for which this program converges. This comparison
demonstrates the validity of our approach and illustrates its
advantage for characterizing clathrate hydrates at temperatures outside
the range accessible to the CSMHYD program.

\subsection{Accuracy of the potential parameters}

The calculations of the relative abundances of a guest species
trapped in clathrate hydrates depend both on the structural
characteristics of the clathrate hydrate under consideration and
on the accuracy of the corresponding interactions between the
trapped molecule and the water cage (Thomas et al., 2008). As a
consequence, the accuracy of the present calculations strongly
depends on the choice of the parameters of the Kihara potential.

In our previous study (Thomas et al., 2008), we have shown that an
appropriate set of parameters for clathrate hydrate studies on Titan has been
given by Parrish and Prausnitz (1972), in which potential and
structural parameters have been self-consistently determined on
experimentally measured clathrate hydrate properties.

To check the accuracy of these parameters in the present
application on Mars, we have compared the relative abundances
obtained with the Kihara parameters given by Sloan (1998) (same as
in the CSMHYD program) to those calculated using the Kihara
parameters given by Parrish and Prausnitz (1972) (Table \ref{Kihara}).
Surprisingly, calculations performed with the parameters from
Parrish and Prausnitz (1972) lead to results at the opposite of those
obtained with the parameters given by Sloan (1998). Indeed, the
parameters from Parrish and Prausnitz (1972) lead to a better trapping
when the clathrate hydrates form at low temperatures than when they form
at high temperatures. This appears very puzzling given that
parameters from Parrish and Prausnitz (1972) and from Sloan (1998) are very
similar, except for the value of the parameter $a$ for CO$_2$ (
the radius of the impenetrable core), which is almost twice larger
in the Sloan's set of parameters than in Parrish and Prausnitz's
parameters. Additional tests performed by varying the $a$ values
in our calculations showed that the relative abundances are in
fact very sensitive to this $a$ parameter, as already stated by
Papadimitriou et al. (2007). Because more recent sets of Kihara
parameters (Jager, 2001; Kang et al., 2001)\footnote{Unfortunately, contrary to the data published by Parrish and Prausnitz (1972), these sets do not provide the complete list of Kihara parameters required by the molecules involved in our system.} give a $a$ value for
CO$_2$ close to that given by Sloan, the value $a_{\rm CO_2}$
given by Parrish and Prausnitz (1972) appears rather suspicious. We have
thus run again our calculations by using the Kihara set of
parameters from Parrish and Prausnitz (1972), in which we have replaced
the suspicious $a_{\rm CO_2}$ value by the one given by Sloan
(1998). The corresponding results are given in Fig.\ref{fig1}
which shows that this combined set of parameters (dashed lines)
give results which are very similar to those obtained with the
whole set of Sloan's parameters for the Kihara potential. However,
because Kihara parameters are only given for a reduced set of
species in Sloan (1998), we finally choose to use here the
parameters given by Parrish and Prausnitz (1972), but with the $a_{\rm
CO_2}$ value given by Sloan (1998).

\section{Variability of the methane trapping in clathrate hydrates}

The statistical approach presented in the previous section has
been used to reinvestigate the work of CC07 and to determine more
accurately the composition of clathrate hydrates formed in the
near subsurface of Mars as a function of temperature and gas phase
composition. In particular, we consider a more plausible
composition of the gas phase by taking into account carbon
dioxyde, nitrogen and argon, namely the most abundant volatiles of
the martian atmosphere\footnote{The present martian atmosphere
contains 95\% of CO$_2$, 2.7\% of N$_2$ and 1.6\% of Ar (Moroz,
1998).}, together with methane in our calculations. Besides, we
determine the composition of clathrate hydrates formed at
temperatures corresponding to the extreme ones measured in the
polar caps (130 K; Kieffer et al., 2001), whereas the work of CC07
was restricted to higher temperatures due to the limitations of
the CSMHYD program.

Six different initial abundances of methane in the gas phase have
been considered in the present study, namely 0.01\%, 0.1\%, 1\%,
10\%, 50\% and 90\%. The largest values are typical of
methane-rich conditions in which CH$_4$ is supplied from below by
microbial or geological processes or from above from ancient
atmospheres. In contrast, the lowest values are more typical of
recent atmospheric compositions. In each case, we assume that the
ratios CO$_2/$N$_2$, CO$_2/$Ar, N$_2/$Ar are equal to those
measured in the present martian atmosphere, and
$\mathnormal{x}_{\rm{CH_4}}+\mathnormal{x}_{\rm{Ar}}+\mathnormal{x}_{\rm{CO}}+\mathnormal{x}_{\rm{N_2}}=1$.

\begin{figure}
\resizebox{\hsize}{!}{\includegraphics[angle=0]{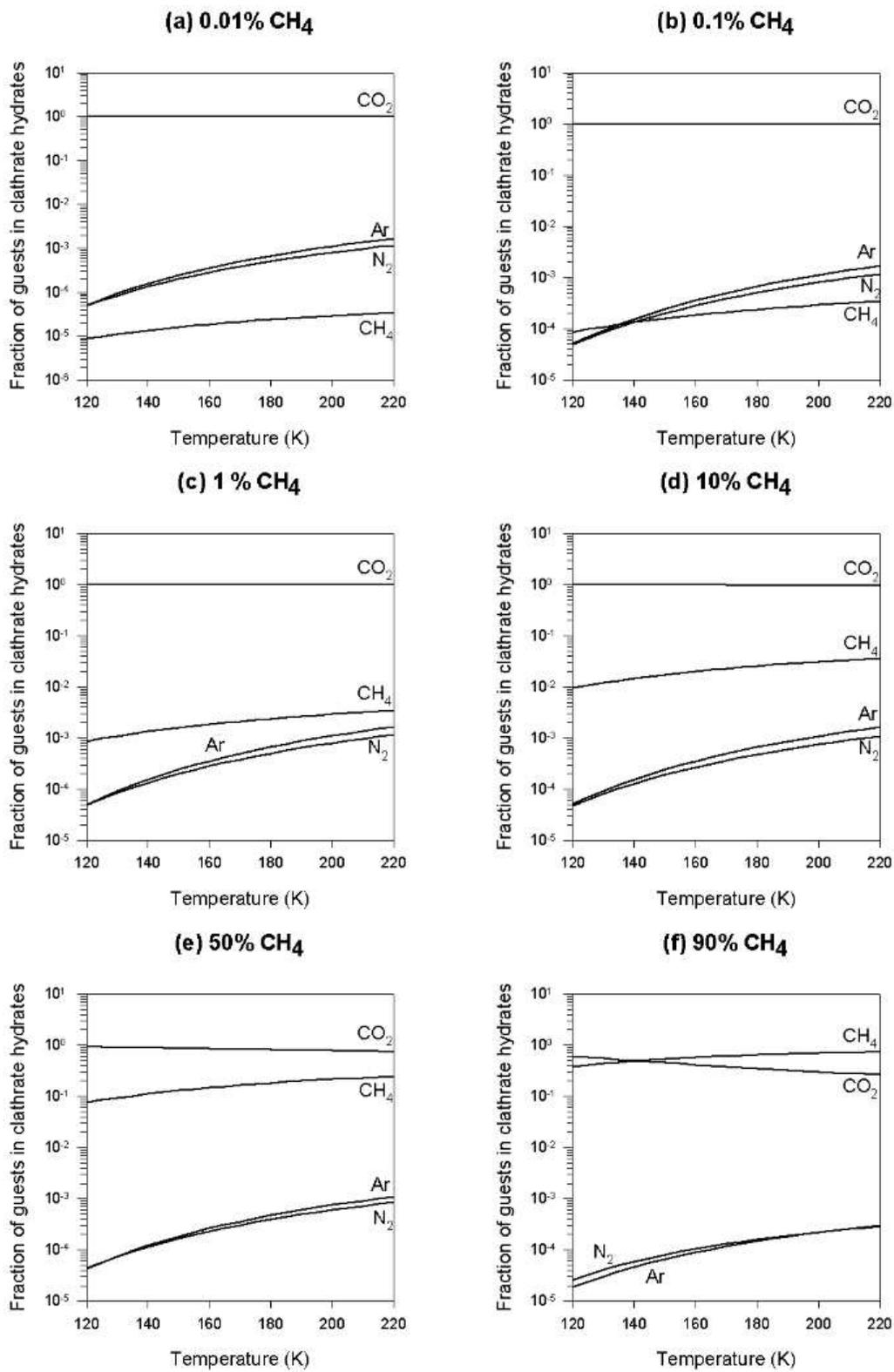}}
%\resizebox{\hsize}{!}{\includegraphics[angle=0]{graphes.pdf}}
\caption{Relative abundances of CO$_2$, CH$_4$, N$_2$ and Ar in
clathrate hydrates as a function of temperature for the different
methane abundances considered in the present work.} \label{PP}
%\caption{Relative abundances of CO$_2$, CH$_4$, N$_2$, CO, O$_2$, Kr, Xe and Ar in
%clathrate hydrates as a function of temperature for the different
%methane abundances considered in the present work.} \label{PP}
\end{figure}

The evolution with temperature of the relative abundances $f_G$
($G = \ $CH$_4$, CO$_2$, N$_2$, Ar) calculated in multiple guest
clathrate hydrates corresponding to the various initial abundances
of CH$_4$ considered here is given in Fig.~\ref{PP}. This figure
shows that the trapping behavior is the same for each situation,
that is the relative abundances of Ar, N$_2$ and CH$_4$ slightly
increase in clathrate hydrates when the formation temperature
increases, whereas that of CO$_2$ slightly decreases, irrespective
of the initial gas phase abundances. Moreover, Ar and N$_2$ are
always poorly trapped in clathrate hydrates whereas the
incorporation of CH$_4$ and CO$_2$ strongly depends on the initial
composition of the gas phase. Indeed, methane is poorly trapped
when its initial abundance in the gas phase is lower than a few
percent (see Figs.~\ref{PP}.a to ~\ref{PP}.d) and, in such a
situation, CO$_2$ fills almost entirely the clathrate hydrates
with a relative abundance $f_{\textrm{CO}_2}\simeq 1$. In
contrast, when the initial gas phase abundance is enriched in
methane, there is a competition between the trapping of CO$_2$ and
that of CH$_4$ which thus have similar abundances in the
corresponding clathrate hydrate (Figs.~\ref{PP}.e to ~\ref{PP}.f).

Another way for characterizing the trapping in clathrate hydrates is to
calculate the abundance ratio for the different gases, which is
defined as the ratio between the relative abundance $f_G$ of a
given gas in the multiple guest clathrate hydrate and its initial gas
phase abundance $x_G$ (Thomas et al., 2007, 2008).
The abundances ratios calculated for CH$_4$, CO$_2$, Ar and N$_2$
in the different situations considered here are given in Table
\ref{results} together with the corresponding relative abundances in the initial gas phase ($x_G$) and in clathrate hydrates ($f_G$). Note
that we have calculated these abundance ratios at the particular
point on the dissociation curves corresponding to the present
average atmospheric pressure on Mars, i.e. $P=7$ mbar.

For this particular point, Table \ref{results} shows that the
abundance ratio of CH$_4$ increases with $x_{CH_4}$, as a
consequence of the larger trapping of methane in the corresponding
multiple guest clathrate hydrates (as indicated by the increase of the
relative abundance $f_{CH_4}$). However, this ratio
remains lower than 1 in all situations, indicating that the
trapping efficiency of CH$_4$ in the multiple guest clathrate hydrates
considered here is quite low. On the contrary, the abundance ratio
of CO$_2$ is always larger than 1, showing the high efficiency of
the trapping of CO$_2$ in clathrate hydrates. Indeed, even when the gas
phase contains less than 10 \% of CO$_2$, the corresponding
multiple guest clathrate hydrate contains about 50 \% of CO$_2$, as
indicated by the values of $f_{CO_2}$ in Table \ref{results}. As a
consequence of this preferential trapping of CO$_2$ with respect
to the other species of the gas phase, the abundance ratio of
CO$_2$ given in Table \ref{results} increases when the initial gas
phase abundance $x_{CO_2}$ decreases. In other words, although the
trapping of CH$_4$ is more and more efficient when the initial gas
phase is enriched in CH$_4$, it remains much less efficient than
the trapping of CO$_2$. Note that Table \ref{results} also shows
that Ar is slightly better trapped than N$_2$. However, the
abundances of these two gases are almost negligible in the
multiple guest clathrate hydrates considered in the present study.

\section{Discussion}

Here, we have revisited the work of CC07 devoted to the
trapping of CH$_4$ in clathrate hydrates that may exist in the
near subsurface of Mars. Our conclusions are similar to those of
CC07, although they are based on a more appropriate set of Kihara
parameters to describe the interactions between water molecules
and guest species. In presence of CO$_2$, a methane-rich
clathrate hydrate can be thermodynamically stable only if the gas
phase is itself strongly enriched in methane. Hence, CH$_4$-rich
clathrate hydrates cannot be formed from the present martian
atmosphere, which has been found to be very poor in methane (Mumma
et al., 2003; Formisano et al., 2004; Krasnopolsky et al., 2004;
Geminale at al., 2008). As a consequence, if they do exist,
CH$_4$-rich clathrate hydrates on Mars should have been formed in
contact with an early martian atmosphere, much richer in CH$_4$
than the present one. For example, planetary impacts by comets and
meteorites in the past could have enriched the martian atmosphere
in CH$_4$ (Kress and McKay, 2004). Also, the detection of gray
crystalline hematite deposits on Mars could be a proof of an early
methane-rich martian atmosphere (Tang et al., 2005). As mentioned
in the Introduction of the present paper, another possible origin
of CH$_4$-rich clathrate hydrate could be a subsurface source.
Indeed, if an internal source of biological (Krasnopolsky et al.,
2004) or geological origin (Oze and Sharma, 2005) has existed (or
still exists) in the martian subsurface, the present calculations
show that stable subsurface CH$_4$-rich clathrate hydrates could
form in contact with the produced methane outgassing towards the
surface.

\section*{Acknowledgements}
This work was supported in part by the French Centre National d'Etudes Spatiales.

\clearpage

\begin{table}[h]
\centering \caption{Two different sets of parameters for the
Kihara potential. $\sigma$ is the Lennard-Jones diameter,
$\epsilon$ is the depth of the potential well, and $a$ is the
radius of the impenetrable core. These parameters derive from
(a) Parrish \& Prausnitz (1972), (b) Sloan (1998).}
\begin{tabular}{clccc}
\hline \hline
Ref             & Molecule   & $\sigma$(\AA)& $ \epsilon/k_B$(K)& $a$(\AA) \\
\hline
(a)                 & CH$_4$     & 3,2398 & 153,17 & 0,300 \\
                            & N$_2$      & 3,2199 & 127,95 & 0,350 \\
                            & CO$_2$     & 2,9681 & 169,09 & 0,360 \\
                            & Ar         & 2,9434 & 170,50 & 0,184 \\
\hline
(b)                 & CH$_4$     & 3,1650 & 154,54 & 0,3834 \\
                            & N$_2$      & 3,0124 & 125,15 & 0,3526 \\
                            & CO$_2$     & 2,9818 & 168,77 & 0,6805 \\
                    & Ar         &   -    &   -    &   -    \\
\hline
\end{tabular}
\label{Kihara}
\end{table}

\clearpage

\begin{table}[h]
\centering \caption{Relative abundances of CH$_4$, CO$_2$, Ar and
N$_2$ in the initial gas phase ($\mathnormal{x}_G$) and in
clathrates ($\mathnormal{f}_G$). These ratios are calculated at
$P=7$ mbar, and at the corresponding temperature on the
dissociation curves.}
\scriptsize{
\begin{tabular}{cccc}
\hline \hline
gaz        & $\mathnormal{x}_G$    & $\mathnormal{f}_G$   & abundance ratio         \\
\hline
CH$_4$     & 1$\times10^{-4}$      & 1.66$\times10^{-5}$  & 0.166                 \\
           & 1$\times10^{-3}$      & 1.66$\times10^{-4}$  & 0.166                     \\
           & 1$\times10^{-2}$      & 1.67$\times10^{-3}$  & 0.167                   \\
           & 0.1                   & 1.77$\times10^{-2}$  & 0.177                   \\
           & 0.5                   & 0.127                  &   0.254                               \\
           & 0.9                   & 0.496                  &   0.551                               \\
\hline
CO$_2$     & 0.957                 & 0.999                  &   1.044                               \\
           & 0.956                 & 0.999                  &   1.045                               \\
           & 0.947                 & 0.998                  &   1.054                               \\
           & 0.861                 & 0.982                  &   1.106                               \\
           & 0.478                 & 0.873                  &   1.826                               \\
           & 0.096                 & 0.504                  &   5.250                               \\
\hline
Ar         & 1.61$\times10^{-2}$   & 2.71$\times10^{-4}$  & 1.68$\times10^{-2}$ \\
           & 1.61$\times10^{-2}$   & 2.71$\times10^{-4}$  & 1.68$\times10^{-2}$ \\
           & 1.59$\times10^{-2}$   & 2.69$\times10^{-4}$  & 1.69$\times10^{-2}$ \\
           & 1.50$\times10^{-2}$   & 2.62$\times10^{-4}$  & 1.75$\times10^{-2}$ \\
           & 0.81$\times10^{-2}$   & 1.69$\times10^{-4}$  & 2.10$\times10^{-2}$ \\
           & 0.16$\times10^{-2}$   & 4.56$\times10^{-5}$  & 2.83$\times10^{-2}$ \\
\hline
N$_2$      & 2.72$\times10^{-2}$   & 2.20$\times10^{-4}$  & 0.81$\times10^{-2}$ \\
           & 2.72$\times10^{-2}$   & 2.20$\times10^{-4}$  & 0.81$\times10^{-2}$ \\
           & 2.69$\times10^{-2}$   & 2.19$\times10^{-4}$  & 0.81$\times10^{-2}$ \\
           & 2.40$\times10^{-2}$   & 2.05$\times10^{-4}$  & 0.85$\times10^{-2}$ \\
           & 1.36$\times10^{-2}$   & 1.54$\times10^{-4}$  & 1.13$\times10^{-2}$ \\
           & 0.27$\times10^{-2}$   & 5.86$\times10^{-5}$  & 2.15$\times10^{-2}$ \\
\hline
\end{tabular}
}
\label{results}
\end{table}

\clearpage

\clearpage

%\clearpage

%\begin{figure}
%\resizebox{\hsize}{!}{\includegraphics[angle=0]{}}
%\caption{.}
%\label{}
%\end{figure}

%\clearpage

\end{document}